\documentclass{article}

\usepackage{algorithm}
\usepackage{algpseudocode}
\usepackage{PRIMEarxiv}
\usepackage{subfig}
\usepackage[utf8]{inputenc} % allow utf-8 input
\usepackage[T1]{fontenc}    % use 8-bit T1 fonts
\usepackage{hyperref}       % hyperlinks
\usepackage{url}            % simple URL typesetting
\usepackage{booktabs}       % professional-quality tables
\usepackage{amsfonts}       % blackboard math symbols
\usepackage{nicefrac}       % compact symbols for 1/2, etc.
\usepackage{microtype}      % microtypography
\usepackage{lipsum}
\usepackage{fancyhdr}       % header
\usepackage{graphicx}       % graphics
\graphicspath{{media/}}     % organize your images and other figures under media/ folder
\usepackage{amsmath}
\usepackage{natbib}
\title{Post-FWI Injection of Learned Priors Using a Flow Matching Model
}

\author{
  Hao Zhang \\
   King Abdullah University of Science and Technology \\
  \texttt{hao.zhang.1@kaust.edu.sa} \\
     \And
Tariq Alkhalifah \\
   King Abdullah University of Science and Technology \\
  \texttt{
tariq.alkhalifah@kaust.edu.sa} \\
}

\begin{document}
\maketitle

\begin{abstract}
Full Waveform Inversion (FWI) is a powerful tool for subsurface velocity reconstruction but remains highly ill-posed, sensitive to acquisition limitations, often requiring some form of regularization to reduce artifacts and enhance resolution. While recent developments have shown that generative models can inject learned priors directly into the FWI optimization process, such approaches typically require additional, computationally expensive inversion iterations. In this study, we propose a post-FWI refinement strategy based on a Flow Matching (FM) generative model, which leverages learned geological priors without re-running FWI. The method guides the deterministic generative process using the FWI result, as well as well logs, if available. Synthetic and field data experiments demonstrate that we can inject well information and our geological expectations (prior) into the provided FWI result, and thus, we can effectively enhance its resolution and geological quality. In fact, the well prior even managed to alter the model depth to fit the well information, which is a form of correcting for depth misties.
\end{abstract}

% keywords can be removed
\section{Introduction}
Full waveform inversion (FWI) is an advanced technique for reconstructing subsurface velocity models by iteratively minimizing the misfit between observed and simulated seismic data \cite{virieux2009overview}. However, FWI is a highly non-unique, ill-posed inverse problem, and its non-uniqueness strongly depends on acquisition geometry and frequency band, as well as the quality of the initial model. As a result, FWI inverted models often contain noticeable artifacts caused by limited illumination, incomplete acquisition coverage, and data noise. Consequently, the inverted velocity models may deviate significantly from geologically realistic structures, particularly in poorly illuminated regions. They also often deviate from the well information, which we consider as ground truth; a phenomenon that we attribute to complex velocity variation and anisotropy \cite{bui2010well,martinez2021well}.

To mitigate these issues, regularization is commonly introduced into the objective function by imposing assumptions on the model, such as sparsity \cite{strong2003edge,guitton2012constrained}, or smoothness \cite{golub1999tikhonov}, or by incorporating prior information, such as well logs\cite{asnaashari2013regularized} and gravity data \cite{blom2017synthetic}, to stabilize the inversion and improve convergence. Nevertheless, these regularization methods typically rely on handcrafted and fixed assumptions of the model, which may introduce bias toward smooth or sparse models, thus reducing resolution and failing to adequately represent complex geological features.

In recent years, various deep learning techniques have been suggested to incorporate prior information into the regularization of FWI  \cite{zhang2019regularized, li2021deep, sun2023FWIlearned, wang2023prior, tan2025FWIdenosing,lewis2017deep, yang2024conditional,he2021reparameterized, dhara2023elastic,wang2026multiscale,zhang2021bayesian,zhang2024bayesian}, as well as other geophysical problems \cite{romero2022plug,wang2026plug}. One representative approach is the plug-and-play (PnP) denoiser \cite{zhang2019regularized, li2021deep, sun2023FWIlearned, wang2023prior, tan2025FWIdenosing}. In this framework, a neural network is pre-trained (either in a supervised or unsupervised manner) to map a noisy velocity model to a clean, high-resolution model with improved structural consistency. During the iterative inversion process, the updated model at each step is refined by applying a denoiser, thereby suppressing noise and enhancing resolution. In addition to explicit denoisers, reparameterization provides an alternative way to impose implicit priors by expressing the model as the output of a neural network. In this framework, optimization can be performed with respect to either the input noise (as in Deep Image Prior \cite{lewis2017deep, yang2024conditional}) or the network parameters\cite{he2021reparameterized, dhara2023elastic,wang2026multiscale}. These approaches are effective because the inductive bias of the neural network constrains the solution space, thereby implicitly regularizing the inversion process. Unlike deterministic methods described above, Bayesian inference \cite{zhang2021bayesian,zhang2024bayesian}, as a probabilistic inversion framework, provides a natural way to incorporate prior information, since the prior is an essential component of the formulation rather than being introduced heuristically. However, similar to classical regularization approaches, these ML-based methods remain tightly coupled with FWI optimization, requiring repeated forward simulations and gradient computations, which leads to significant computational cost.

In many practical scenarios, FWI is often performed without explicit regularization, or with only limited regularization such as fixed smoothness constraints, which may produce  geologically implausible velocity models. Furthermore, incorporating existing regularization techniques directly into FWI often requires repeated forward modeling and gradient calculations, significantly increasing computational cost. This motivates the following question: Can we inject prior information into the FWI result without performing additional expensive FWI iterations? In particular, can learned priors be incorporated after the FWI process has converged, serving as a post-processing step that refines the inverted model?

This idea is not entirely new. Injecting learned priors directly into models obtained from traditional processing methods has attracted increasing attention over the years. For example,  Li et al.\cite{li2023self} employed a vision transformer (ViT) to learn a mapping from seismic images to high-resolution facies models. Similarly, Brandolin et al.\cite{brandolin2026velocity} guided the generation of high-resolution velocity models by incorporating well-log data and structural information extracted from the initial model. These approaches demonstrate the potential of post-processing strategies for incorporating geological priors. However, they typically rely on direct supervised mappings or hand-crafted guidance. 

In this study, we propose a post-FWI refinement strategy that injects learned geological priors into the FWI results using an unconditional flow matching (FM) model \cite{lipman2022flow,liu2022flow,esser2024scaling}, without explicitly enforcing data-fitting constraints during generation. FM is a continuous-time generative modeling framework that learns deterministic transport between data distributions, enabling stable training and efficient sampling. Diffusion models, as a prominent class of continuous-time generative models, have attracted significant attention in seismic data processing \cite{zhu2023diffusion,zhang2024conditional,li2024conditional,meng2024posterior} and velocity model building \cite{wang2024geological,zhang2024diffusionvel,zhang2025multi,zhanger2025} due to their strong ability to learn data priors. The key distinction is that diffusion models learn the score function and induce the probability flow implicitly, whereas FM directly learns the corresponding diffusion flow velocity field. Compared to conventional Diffusion models, FMs are more stable in training and faster in sampling\cite{esser2024scaling,ma2024sit,schusterbauer2025diff2flow}. and these advantages make them particularly well suited for post-inversion refinement guided by FWI results.

The objective of our method is to generate velocity models that remain consistent with structural features provided by the FWI results and, when available, well-log observations, while simultaneously preserving the geological priors learned by the FM model. To achieve this, we guide the generation process using observational constraints by directly optimizing the latent variables along the generation trajectory. Since the FM model is trained on high-resolution velocity models that contain richer geological details than those typically recovered by FWI, a smoothing operator is introduced when applying FWI guidance to preserve the high-frequency structures learned by the generative prior. For well-log guidance, a masking operator is applied, since well-log observations are only available at sparse spatial locations. Unlike reconstruction-based \cite{wang2024controllable} or classifier-guided \cite{taufik2025diffusion} approaches that require explicit propagation of sparse well information to global structures, the proposed method directly constrains the observed locations while relying on the learned generative prior to maintain global geological consistency. This design allows the generated model to preserve the large-scale structures from FWI while incorporating localized high-resolution constraints from well logs within the learned geological prior. We validate the proposed method on both synthetic and field datasets and demonstrate its effectiveness in improving subsurface velocity models through post-inversion prior injection.

The main contributions of this study are summarized as follows:

\begin{itemize}
    \item We propose a post-FWI prior injection framework that refines the final FWI velocity models using an FM generative model, without requiring additional FWI iterations or explicit data-fitting during generation.

    \item We introduce a guidance-based generative refinement strategy that incorporates both large-scale structural information from FWI results and localized high-resolution constraints from well logs within a unified generative framework.

    \item To preserve the high-frequency geological details learned by the generative prior, we design a smoothed FWI-guidance operator that constrains only the large-scale structures during generation, avoiding over-smoothing and loss of prior-consistent details. 
    
    \item We propose a sparse well-log guidance mechanism that directly injects localized well observations into the generative process through masking-based constraints, avoiding the need for classifier guidance or explicit reconstruction-based extrapolation to global structures.

    \item We validate the proposed method on both synthetic and field datasets, demonstrating improved geological consistency, fault recovery, and structural continuity compared with conventional FWI results.
\end{itemize}

\section{Related work}

The approach suggested here is closely related to the concept of proximal solvers, which have lately become important in modern optimization, particularly for handling non-smooth regularization terms that arise in inverse problems. Instead of directly minimizing the data-fitting and regularization terms of the objective, these methods rely on iterative schemes that alternate between gradient-based updates for smooth data misfit terms and proximal mappings that implicitly enforce regularization. This framework enables the incorporation of a wide range of priors, including sparsity-promoting norms, total variation, and constraint-based formulations, while maintaining computational efficiency \cite{parikh2014, beck2009}.

For FWI, proximal solvers provide a powerful mechanism to mitigate the inherent ill-posedness and nonlinearity of the problem. By decoupling the data-fitting term from the regularization, we can integrate advanced priors into FWI, including salt flooding, sparsity in transformed domains, and more recently, learned priors derived from neural networks and generative models \cite{wang2023prior}. Proximal splitting techniques, such as the split Bregman \cite{10.1190/geo2018-0146.1} and the alternating direction method of multipliers (ADMM), have been used to incorporate total variation and other structural constraints into FWI workflows \cite{FU2020104030}, improving stability and convergence in the presence of noise and limited acquisition geometries \cite{boyd2011}. Furthermore, the proximal framework naturally extends to PnP and regularization-by-denoising (RED) approaches, where sophisticated denoisers or diffusion models act as implicit priors, offering new opportunities for combining physics-based inversion with data-driven regularization \cite{6737048,https://doi.org/10.1029/2024JH000125}.

In this study, we utilize the concept of decoupling the data fitting and regularization within the framework of post processing or refining the FWI result. This refinement is executed using a trained FM model guided by the inverted velocity model, as well as any velocity information from a well if available.

\par
\vspace{0.4cm}
\section{Theory}

In this section, we first review the role of regularization in conventional FWI, and the ability to decouple it from the main data fitting term. We follow that by introducing FM as an unconditional generation process that stores, through training on samples, the prior distribution.  Then, guidance of the generation process using data (in our case, the FWI result and possibly well information) is introduced to obtain a sample from the posterior, which we use to formulate our post FWI prior injection methodology.

\subsection{Conventional FWI with Regularization}

FWI aims to recover the subsurface velocity model from an initial estimate by iteratively minimizing the misfit between observed and simulated seismic data. The objective function is typically formulated as
\begin{equation}
\mathbf{m}^* = \arg\min_{\mathbf{m}} \left\{
\mathcal{L}(\mathbf{F}(\mathbf{m}), d_{\text{obs}})
+ \lambda \mathcal{R}(\mathbf{m})
\right\}.
\label{eq:1}
\end{equation}
Here, $\mathcal{L}$ denotes the data misfit term, $\mathbf{m}$ represents the parameters of the model (e.g., velocity), $\mathbf{F}$ is the forward modeling operator, and thus, $\mathbf{F}(\mathbf{m})$ denotes the synthetic data corresponding to model $\mathbf{m}$, $d_{\text{obs}}$ represents the observed seismic data, $\mathcal{R}$ is the regularization term, and $\lambda$ is the regularization weight. Using the squared $\ell_2$ norm, the data misfit term can be written as
\begin{equation}
\mathcal{L}(\mathbf{F}(\mathbf{m}), d_{\text{obs}})
= \frac{1}{2} \left\| \mathbf{F}(\mathbf{m}) - d_{\text{obs}} \right\|_2^2.
\label{eq:2}
\end{equation}

Synthetic data $\mathbf{F}(\mathbf{m})$ are obtained by solving the acoustic wave equation:
\begin{equation}
\frac{1}{v^2(\mathbf{x})} \frac{\partial^2 u(\mathbf{x}, t)}{\partial t^2}
- \nabla^2 u(\mathbf{x}, t)
= s(\mathbf{x}, t),
\label{eq:3}
\end{equation}
where $u(\mathbf{x}, t)$ is the wavefield as a function of space coordinates $\mathbf{x}$ and time coordinate $t$, $v(\mathbf{x})$ is the velocity field, and $s(\mathbf{x}, t)$ denotes the source term. The synthetic data corresponds to sampling the wavefield, $u$, at receiver locations.

FWI is a highly nonlinear and ill-posed inverse problem. The regularization term $\mathcal{R}(\mathbf{m})$ incorporates prior information about the model and stabilizes the inversion. Depending on its formulation, $\mathcal{R}(\mathbf{m})$ can promote smoothness (e.g., $\|\nabla \mathbf{m}\|_2^2$), sparsity (e.g., $\|\nabla \mathbf{m}\|_1$), or other structural constraints, thereby mitigating ill-posedness and reducing sensitivity to noise and incomplete data.

FWI proceeds by iteratively updating the model parameters using a gradient-based optimization. At iteration $k$, the update is given by
\begin{equation}
\mathbf{m}^{(k+1)} = \mathbf{m}^{(k)} - \alpha^{(k)} \nabla_{\mathbf{m}} \left[
\mathcal{L}(\mathbf{F}(\mathbf{m}^{(k)}), d_{\text{obs}}) + \lambda \mathcal{R}(\mathbf{m}^{(k)})
\right],
\label{eq:4}
\end{equation}
where $\alpha^{(k)}$ is the step size, and the gradient is typically calculated using the adjoint-state method.

From Equation \ref{eq:1}, the regularization term is conventionally incorporated directly into the FWI, and therefore its influence is incorporated within the FWI iterations. This coupling requires repeated forward simulations and gradient evaluations, leading to a significant computational cost. In practice, this becomes particularly challenging when only the final FWI results are available (e.g., from external sources), and the original data or computational resources required to re-run the inversion are inaccessible. To address this limitation, this paper attempts to project prior information directly onto FWI results without requiring iterative inversion.

\subsection{Decoupling the regularization}

A key advantage of proximal solvers lies in their ability to decouple the data-fitting term from the regularization in composite optimization problems.
Consider our optimization problem in equation~\ref{eq:1}, rather than treating it as a single problem, proximal methods divide the update into two steps: a gradient descent step on the first term and a proximal mapping associated with $\mathcal{R}(\mathbf{m})$.
We introduce an auxiliary variable $\mathbf{v} = \mathbf{m}$, which will allow the problem to be formulated as a constrained optimization:
\begin{equation}
\mathbf{m}^* = \arg\min_{\mathbf{m}} \left\{
\mathcal{L}(\mathbf{F}(\mathbf{m}), d_{\text{obs}})
+ \lambda \mathcal{R}(\mathbf{v})
\right\} \quad \text{subject to} \quad \mathbf{v} = \mathbf{m}.
\label{eq:1b}
\end{equation}
which can be solved via alternating updates that involve independent proximal evaluations of $\mathcal{L}(\mathbf{F}(\mathbf{m}), d_{\text{obs}})$ and $\mathcal{R}(\mathbf{v})$. This separation allows each component to be handled using specialized solvers \cite{parikh2014}.

Importantly, this decoupling has enabled the integration of advanced data-driven priors through PnP and RED frameworks, where the proximal operator is replaced by a denoiser or generative model. In this setting, the inversion step enforces data consistency, while the proximal (denoising) step projects the solution onto a learned manifold of plausible models \cite{venkatakrishnan2013plug,romano2017little}. This paradigm provides an opportunity to fully decouple the regularization step as a separate post-FWI processing step in which we inject priors learned by a generative model, and to implement the equivalent proximal solver we use FM.

\begin{figure*}[htbp!]
    \centering
      {\includegraphics[width=1\columnwidth]{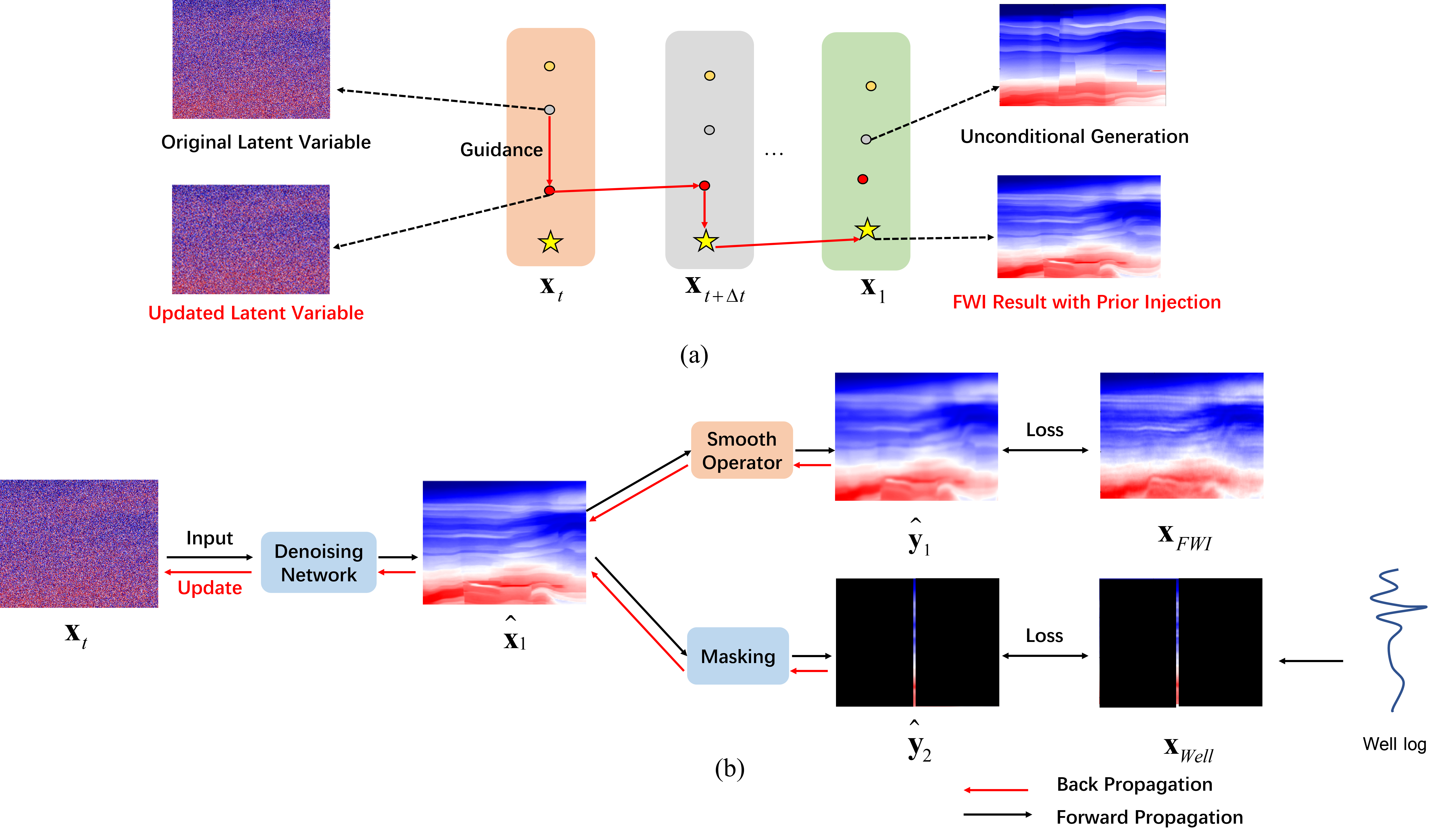}}
      \caption{A graphical illustration of the proposed method: (a) Guided generation; (b) Latent optimization. In (a), given a latent $\mathbf{x}_t$, deterministic flow yields a unique unconditional sample. Guidance perturbs the trajectory across timesteps to steer generation toward the target (e.g., FWI with injected priors), as illustrated by yellow stars in (a). In (b), $\mathbf{x}_t$ is passed through a denoiser (U-Net/DiT) to predict the velocity field (Equation \ref{eq:6}), from which the clean image is obtained (Equation \ref{eq:10}). A forward operator generates synthetic data for comparison with observations. FWI guidance uses a smoothing operator, while well-log guidance uses masking. The weighted loss is backpropagated through the forward operator and denoiser to update $\mathbf{x}_t$.}
    \label{fig:overall_workflow}
\end{figure*}

\subsection{Flow Matching}
\subsubsection{Unconditional Generation}
Given a target data (velocity models) distribution $\mathbf{x}_1 \sim p_{\mathrm{data}}$ and a simple latent distribution
$\mathbf{x}_0 \sim p_0$ (typically a Gaussian), FM learns a deterministic
transformation between the two distributions by modeling the ODE
\begin{equation}
\frac{d \mathbf{x}_t}{d t} = \mathbf{v}_\theta(\mathbf{x}_t, t), \qquad t \in [0,1],
\label{eq:5}
\end{equation}
where $\mathbf{v}_\theta(\mathbf{x}_t,t)$ is a time-dependent velocity field parameterized by a neural network. The model is trained by minimizing a supervised objective that matches the predicted velocity
to a prescribed target velocity $\mathbf{v}^\ast(\mathbf{x}_t, t)$,
\begin{equation}
\mathcal{L}_{\mathrm{FM}}(\theta)
=
\mathbb{E}_{\mathbf{x}_t, t}
\left[
\left\|
\mathbf{v}_\theta(\mathbf{x}_t, t)
-
\mathbf{v}^\ast(\mathbf{x}_t, t)
\right\|_2^2
\right].
\label{eq:6}
\end{equation}

From Equation~\ref{eq:5}, we observe that once $\mathbf{x}_t$ is defined, the corresponding true velocity is uniquely determined. Here, we adopt the widely used parameterization of $\mathbf{x}_t$ as
\begin{equation}
\mathbf{x}_t = (1 - t)\mathbf{x}_0 + t\,\mathbf{x}_1,
\label{eq:7}
\end{equation}
which corresponds to a linear interpolation between $\mathbf{x}_0$ and $\mathbf{x}_1$. 
This leads to  the true constant velocity vector:
\begin{equation}
\mathbf{v}^\ast = \mathbf{x}_0 - \mathbf{x}_1.
\label{eq:8}
\end{equation}

After training, samples are generated deterministically by integrating the learned velocity field along the trajectory starting from an initial latent variable $\mathbf{x}_{t}$ sampled using Equation \ref{eq:7},
\begin{equation}
\mathbf{x}_{t+\Delta t} = \mathbf{x}_t + \mathbf{v}_\theta(\mathbf{x}_t, t)\,\Delta t .
\label{eq:9}
\end{equation}

Once trained, the unconditional FM model learns and stores a rich prior distribution over subsurface models and enables efficient sampling from this distribution.

\begin{algorithm}
\caption{Latent Variable Optimization for Post-FWI Processing}
\begin{algorithmic}[1]
\Require Pre-trained FM model $\mathbf{v}_{\theta}$, total number of time steps $T$, 
FWI result $\mathbf{x}_{\mathrm{FWI}}$, well-log embedded image $\mathbf{x}_{\mathrm{well}}$,  starting time step $t_s$, smoothing operator $\mathcal{F}_1(\cdot)$, masking operator $\mathcal{F}_2(\cdot)$, noise sample $\mathbf{z} \sim \mathcal{N}(0,I)$,
 number of optimization steps $N$ and loss function $\mathcal{L}$
\Ensure Post-FWI prior injected result $\mathbf{x}_1$

\State $\Delta t \gets 1/T$ \Comment{Time discretization step}
\State $\mathbf{x}_{t_s} \gets t_s\Delta t\, \mathbf{x}_{\mathrm{FWI}} + (1-t_s\Delta t)\, \mathbf{z}$
 \Comment{Initialize the starting noise}

\For{$s = t_s$ up to $T-1$}
 \State  $t =s\Delta t$
    \If{Guidance}
    \For{$k = 1$ to $N$}
        \State $\hat{\mathbf{x}}_1 \gets \mathbf{x}_t +(1- t)\, \mathbf{v}_\theta(\mathbf{x}_t, t)$
 \Comment{Predict the clean image}
        \State $\mathcal{L} \gets \mathcal{L}(\mathcal{F}_1(\hat{\mathbf{x}}_1), \mathbf{x}_{\mathrm{FWI}})+\mathcal{L}(\mathcal{F}_2(\hat{\mathbf{x}}_1), \mathbf{x}_{\mathrm{Well}})$
       \Comment{Compute the loss}
        \State $\mathbf{x}_t \gets \mathrm{Optimize}(\mathbf{x}_t, \mathcal{L})$
    \Comment{Optimize the latent variable}
   \EndFor
   \EndIf
    \State $\mathbf{x}_{t+\Delta t} = \mathbf{x}_t + \mathbf{v}_\theta(\mathbf{x}_t, t)\,\Delta t $
\EndFor

\State \Return ${\mathbf{x}}_1$
\label{alg:2}
\end{algorithmic}
\end{algorithm}

\subsubsection{Guided Generation}

As shown in Figure~\ref{fig:overall_workflow} (a)., the FM generation process is deterministic, meaning that each starting latent variable $\mathbf{x}_{t_s}$ (sampled using Equation \ref{eq:7}) corresponds to a unique generation trajectory and therefore a unique generated sample within the learned prior distribution. Consequently, starting from randomly sampled $x_{t_s}$ typically produces random samples from the learned prior. For a desired target model, there theoretically exists a corresponding trajectory (or equivalently, an $x_{t_s}$) capable of generating it. We can initialize the generation process with the sampled $x_{t_s}$ but progressively steer the trajectory toward the desired posterior solution through guidance. During generation, the latent variable \(x_t\) is treated as an optimization variable and iteratively updated so that the generated model remains consistent with the given observations (e.g., FWI results or well logs). The detailed latent optimization procedure is illustrated in Figure~\ref{fig:overall_workflow} (b). Given the current latent state $\mathbf{x}_t$, the predicted clean sample is first estimated as
\begin{equation}
\hat{\mathbf{x}}_1 = \mathbf{x}_t+ (1- t)\, \mathbf{v}_\theta(\mathbf{x}_t, t).
\label{eq:10}
\end{equation}

Then, a forward operator $\mathcal{F}(\cdot)$ is applied to $\hat{\mathbf{x}}_1$ to produce a
synthetic observation $\hat{\mathbf{y}} = \mathcal{F}(\hat{\mathbf{x}}_1)$, which is compared
with the observed data(The inverted FWI velocity model or, if available, well log) $\mathbf{y}$ through the data-consistency loss
\begin{equation}
\mathcal{L}(\mathbf{x}_t) =
\frac{1}{2}
\left\|
\hat{\mathbf{y}}(\mathbf{x}_t) - \mathbf{y}
\right\|_2^2 .
\label{eq:11}
\end{equation}
The latent variable is updated by gradient descent,
\begin{equation}
\mathbf{x}_t \leftarrow \mathbf{x}_t - \alpha \, \nabla_{\mathbf{x}_t} \mathcal{L},
\label{eq:12}
\end{equation}
where the gradient is obtained using the chain rule and Equation \ref{eq:10}:
\begin{equation}
\nabla_{\mathbf{x}_t} \mathcal{L}
=
\frac{\partial \mathcal{L}}{\partial \hat{\mathbf{x}}_1}
\left(
1 -
(1-t) \frac{\partial v_\theta(\mathbf{x}_t, t)}{\partial \mathbf{x}_t}
\right),
\label{eq:13}
\end{equation}
and $\alpha$ controls the strength of the guidance. 

% \begin{figure}[htbp!]
%     \centering
%       {\includegraphics[width=1\columnwidth]{Fig/training_datasets_all.png}}
     
% \caption{Samples of the training dataset. (a) CGG priors. (b) Otway priors}
% \label{fig:training datasets}
% \end{figure}

\subsection{Prior Injection via Guidance}
In this study, we incorporate two types of guidance: FWI-based guidance and well-log-based guidance, to inject their information into the generative process. As illustrated in Figure~\ref{fig:overall_workflow} (b), the FWI-based guidance is derived from a precomputed FWI result and is used without additional data fitting, primarily constraining large-scale velocity and structural features. In contrast, the well-log-based guidance introduces high-resolution local information to refine the model in regions where measurements are available. These two sources of information are coupled through their respective forward operators, enabling a unified integration within the generative framework. The overall post-FWI prior injection workflow is summarized in Algorithm~\ref{alg:2}.

\subsubsection{FWI Guidance}

The key of the FWI-based guidance is the forward operator that maps the generated model to the FWI result. In practice, designing such an operator is challenging. The FWI result is typically limited in resolution due to the band-limited nature of the observed data and is often contaminated by artifacts, which are difficult to model using a simple linear operator. Here, we approximate the forward operator $\mathcal{F}_1(\cdot)$ by a Gaussian smoothing function, which captures the dominant resolution characteristics of the FWI result while avoiding explicit modeling of complex artifacts.

Specifically, we define
\begin{equation}
\mathcal{F}_1(\hat{\mathbf{x}}_1) = G_\sigma * \hat{\mathbf{x}}_1,
\label{eq:14}
\end{equation}
where $*$ denotes convolution and $G_\sigma$ is a Gaussian kernel given by
\begin{equation}
G_\sigma(\mathbf{x}) = \frac{1}{(2\pi\sigma^2)^{d/2}} \exp\left(-\frac{\|\mathbf{x}\|^2}{2\sigma^2}\right),
\label{eq:15}
\end{equation}
where $\sigma$ controls the smoothing scale and $d$ denotes the spatial dimension. In practice, $\sigma$ can be adjusted according to the resolution of FWI results. 

\subsubsection{Well-Log Guidance}
Well-log constraints can be incorporated through three startegies: classifier guidance, conditional FM, and reconstructive guidance. However, these methods have notable limitations. Classifier guidance and conditional FM require additional training, increasing computational cost, and reducing model flexibility. In contrast, reconstructive guidance avoids retraining, but relies on extrapolating sparse well-log information into global constraints, which may introduce bias and lead to unstable guidance.

Instead, we propose a simple and efficient masking-based guidance strategy. In this case, we define a masking operator as the forward operator $\mathcal{F}_2(\cdot)$ to align the generated model with the well logs. This masking operator extracts the velocity profile at the well locations from the generated model.
\begin{equation}
\mathcal{F}_2(\hat{\mathbf{x}}_1) = M \odot \hat{\mathbf{x}}_1,
\label{eq:16}
\end{equation}
where $\odot$ denotes element-wise multiplication and $M$ is a binary mask with the same dimension as $m$, defined as
\begin{equation}
M(\mathbf{x})=
\left\{
\begin{array}{ll}
1, & \mathbf{x}\in\Omega_{\mathrm{well}},\\
0, & \text{otherwise},
\end{array}
\right.
\label{eq:17}
\end{equation}
where $\Omega_{\text{well}}$ denotes the set of spatial locations that correspond to the trajectory of the well. The extracted velocity profile $\mathcal{F}_2(\hat{\mathbf{x}}_1)$ is directly compared with the observed well log \(m_{\text{well}}\) to compute the loss
\begin{equation}
\mathcal{L}_{\text{well}} = \left\| \mathcal{F}_2(\hat{\mathbf{x}}_1) - m_{\text{well}} \right\|^2.
\label{eq:18}
\end{equation}

Since the well-log observations are only available at sparse spatial locations, the corresponding gradients are nonzero only along the well locations. In conventional reconstruction-based guidance approaches, additional spatial propagation or extrapolation strategies are often required to extend these sparse constraints to global structures. In contrast, the proposed method directly constrains the observed locations while relying on the learned generative prior to propagate the influence of the sparse observations through the generation process, thereby maintaining global geological consistency without explicit gradient extrapolation. We will show the details in the CGG dataset example (Figure \ref{fig:gradient_comparasion}). 

\subsubsection{Joint Constraint}
We can easily guide the generation using both the FWI result and the well log by combining these two constraints into the loss function:
\begin{equation}
\mathcal{L} =
\lambda_1 \left\| \mathcal{F}_1(\hat{\mathbf{x}}) -{\mathbf{x}}_{\text{FWI}} \right\|_2^2
+
\lambda_2 \left\| M \odot \hat{\mathbf{x}}-  {\mathbf{x}}_{\text{well}}\right\|_2^2.
\label{eq:19}
\end{equation}

Here, $\lambda_1$ and $\lambda_2$ balance the contributions of the two constraints. This formulation allows the FWI result to constrain the large-scale background velocity and structure, while the well log provides high-resolution local information. However, in field data, significant discrepancies often exist between the FWI-derived velocity and well log measurements. As a result, directly enforcing this joint constraint can lead to poor convergence and degraded inversion results (see Figure~\ref{fig:cgg_welllog_results} (a)). To address this issue, we further propose a frequency-decomposed constraint, where the FWI result constrains the low-frequency components, while only the high-frequency components of the well log are used for guidance. The resulting formulation is given by
\begin{equation}
\mathcal{L} =
\lambda_1 \left\| \mathcal{F}_1(\hat{\mathbf{x}}) - \mathbf{x}_{\text{FWI}} \right\|_2^2
+
\lambda_2 \left\| M \odot \mathcal{F}_3(\hat{\mathbf{x}}) -\mathcal{F}_3({\mathbf{x}}_{\text{well}}) \right\|_2^2,
\label{eq:20}
\end{equation}
where $\mathcal{F}_3 $ is a high-pass filter. Using Equation \ref{eq:20}, the large-scale structures provided by FWI are preserved, while the high-frequency information from the well logs is progressively injected into the generated model.
%Bibliography

\section{Results}

\begin{figure}[htbp!]
    \centering
      {\includegraphics[width=1\columnwidth]{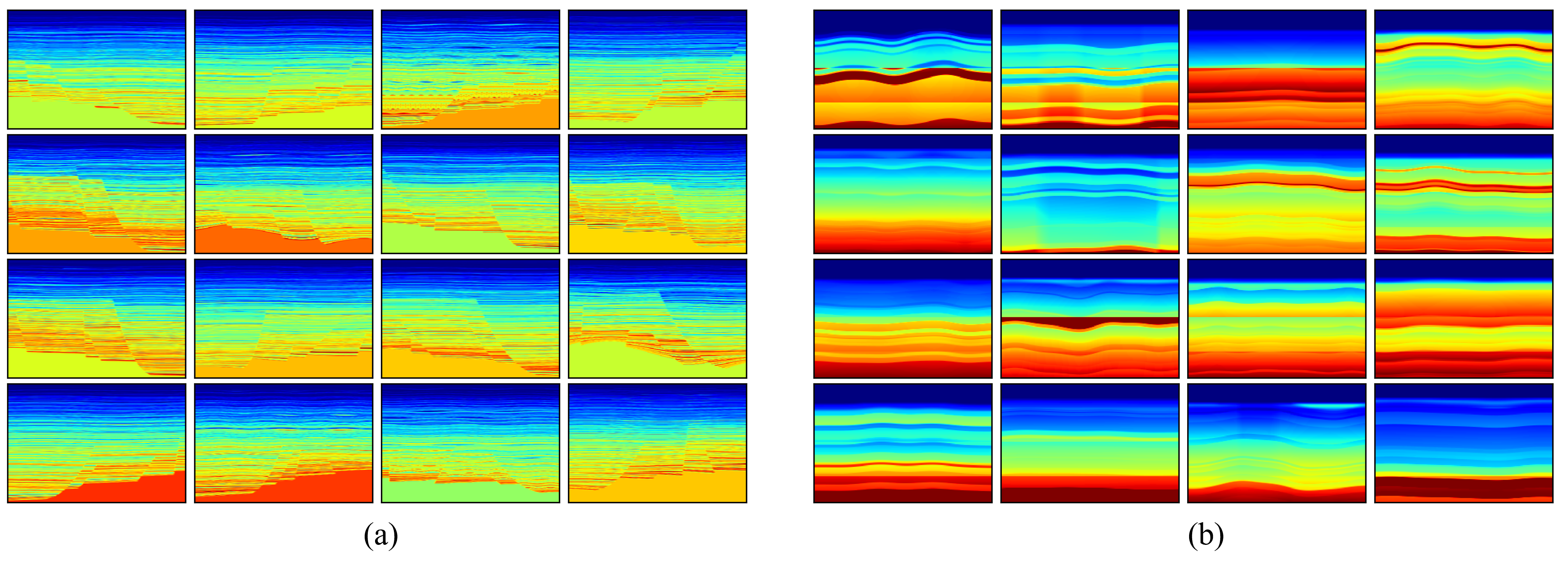}}
     
\caption{Samples of the training dataset. (a) Otway priors. (b) CGG priors.}
\label{fig:training datasets}
\end{figure}

\subsection{Datasets and Training}
To learn the priors used for injection, we first train the FM model on geological prior datasets. Two representative geological prior datasets are prepared: Otway priors (extracted from the Otway model which belongs to Stage 2C of the Otway project by CO2CRC Limited \cite{glubokovskikh2016seismic}, Figure~\ref{fig:training datasets} (a)) and CGG priors (we refer to it CGG because it reflects what we expect the Earth model corresponding to the CGG data would look like, Figure~\ref{fig:training datasets} (b)). The Otway dataset is characterized by thin stratified layers with fault structures, whereas the CGG dataset contains thicker and more continuous geological layers. Each dataset contains 5{,}000 samples with a spatial resolution of \(320 \times 304\).

\begin{figure}[htbp!]
    \centering
      {\includegraphics[width=1\columnwidth]{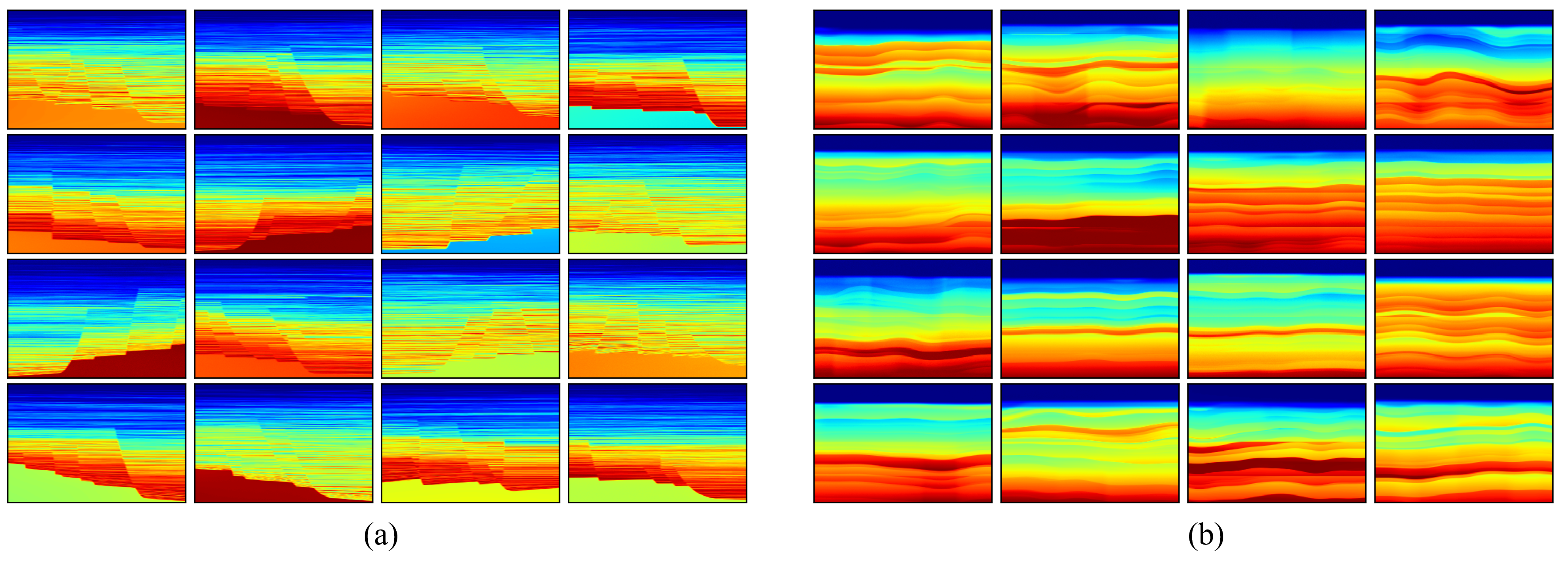}}
     
\caption{Generated samples using the trained class-label FM model. (a) class of Otway priors. (b) class of  CGG priors.}
\label{fig:unconditional_generation}
\end{figure}

We adopt a class-conditional FM framework, where the class label specifies the prior type during both training and sampling. This enables a single model to learn multiple prior distributions while maintaining controllability over the generated velocity models. The FM network is based on a U-Net architecture adapted from~\cite{rombach2022high}. The model is trained directly in the velocity-model space rather than in a compressed latent space, and optimized using the objective defined in Equation~\ref{eq:6}. Training is performed using the AdamW optimizer with a batch size of 4 and an initial learning rate of \(5\times10^{-5}\) for 110 epochs. Representative unconditional samples generated by the trained FM model are shown in Figure~\ref{fig:unconditional_generation}, demonstrating that the model successfully captures the structural characteristics of the corresponding geological priors.

\subsection{Synthetic Example: Otway}
\begin{figure}[h!]
    \centering
      {\includegraphics[width=1\columnwidth]{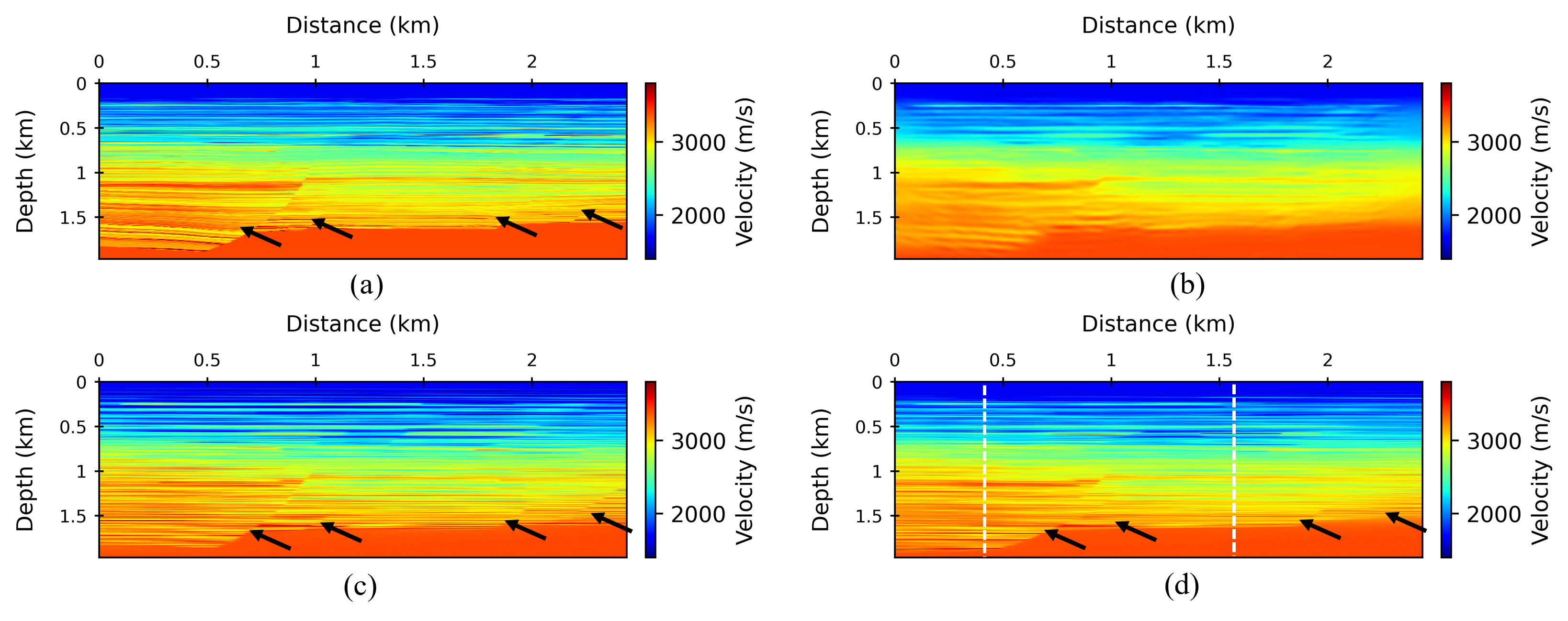}}
\caption{Prior injections for the Otway synthetic dataset. (a) True velocity. (b) FWI result. (c) The refined version after injecting the FM-stored prior. (d) The refined version after injecting the well log data, as well, where the two dashed lines indicate the location of the wells. The black arrows point to the fault locations that should be recovered. }
\label{fig:otway_results}
\end{figure}

\begin{figure}[h!]
    \centering
      {\includegraphics[width=0.97\columnwidth]{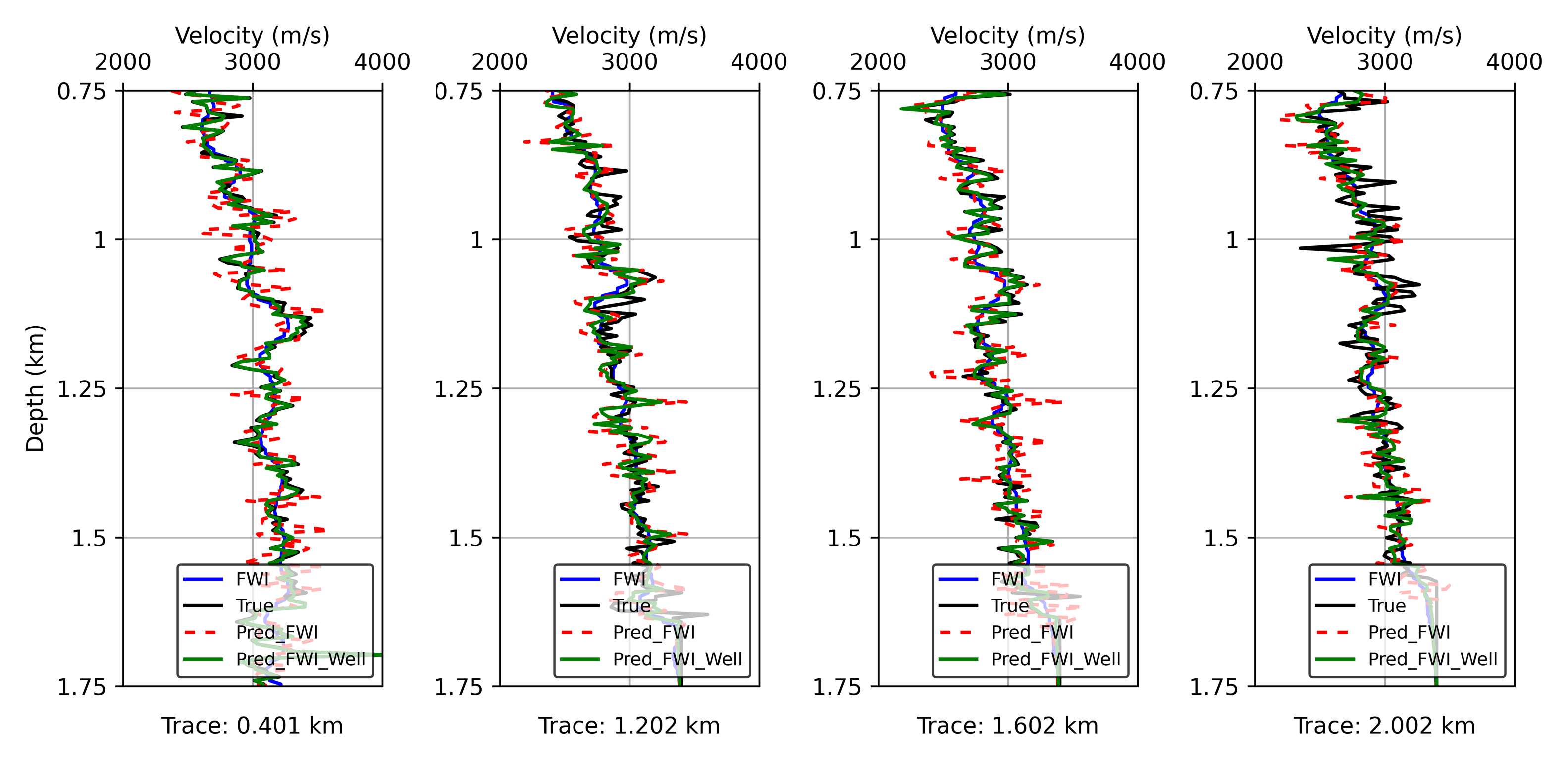}}
    
\caption{Comparison of velocity profiles of the predicted results for the Otway synthetic example at four locations. Profiles at 0.401 km and 1.602 km are located near the well, whereas profiles at 1.202 km and 2.002 km are located away from the well.}

\label{fig:otway_welllogs_comparasion}
\end{figure}

We first evaluate the proposed method using the Otway synthetic dataset. Following the implementation described in Algorithm~1, the total number of generation timesteps is set to \(T=100\), while the reverse process is initiated from an intermediate timestep \(t_s=50\), such that only the final 50 reverse-generation steps are executed to reduce computational cost. At each generation timestep, one latent optimization step (\(N=1\)) is performed to enforce consistency with the given constraints. The original Otway velocity model was resampled such that the velocity models were resized to \(304\) lateral samples and \(320\) vertical samples, consistent with the spatial resolution used during training.  Figure \ref{fig:otway_results} (a) and (b) show the true velocity model and the corresponding FWI inverted result, respectively. Figure \ref{fig:otway_results} (c) shows the predicted sample using the learned Otway priors under the guidance of the FWI result in Figure \ref{fig:otway_results} (b). Compared with the original FWI result, inversion artifacts are effectively suppressed, and the model resolution is enhanced due to the incorporation of such learned priors. As highlighted by the arrows, the faults have been clearly recovered. Building on this, Figure \ref{fig:otway_results} (d) presents the prediction further guided by the well log data using Equation. \ref{eq:19}. The velocity accuracy is evidently improved, particularly near the well locations. This improvement is further supported by the detailed profile comparisons shown in Figure \ref{fig:otway_welllogs_comparasion}. However, the faults are not as clearly resolved as in the result guided solely by FWI. This observation suggests a possible trade-off between preserving structural information from the FWI result and incorporating constraints from well log data. Overall, these results demonstrate that, with appropriate priors, the proposed method can effectively project meaningful geological structures onto the FWI result while improving local accuracy.

\begin{figure}[h!]
    \centering
      {\includegraphics[width=0.8\columnwidth]{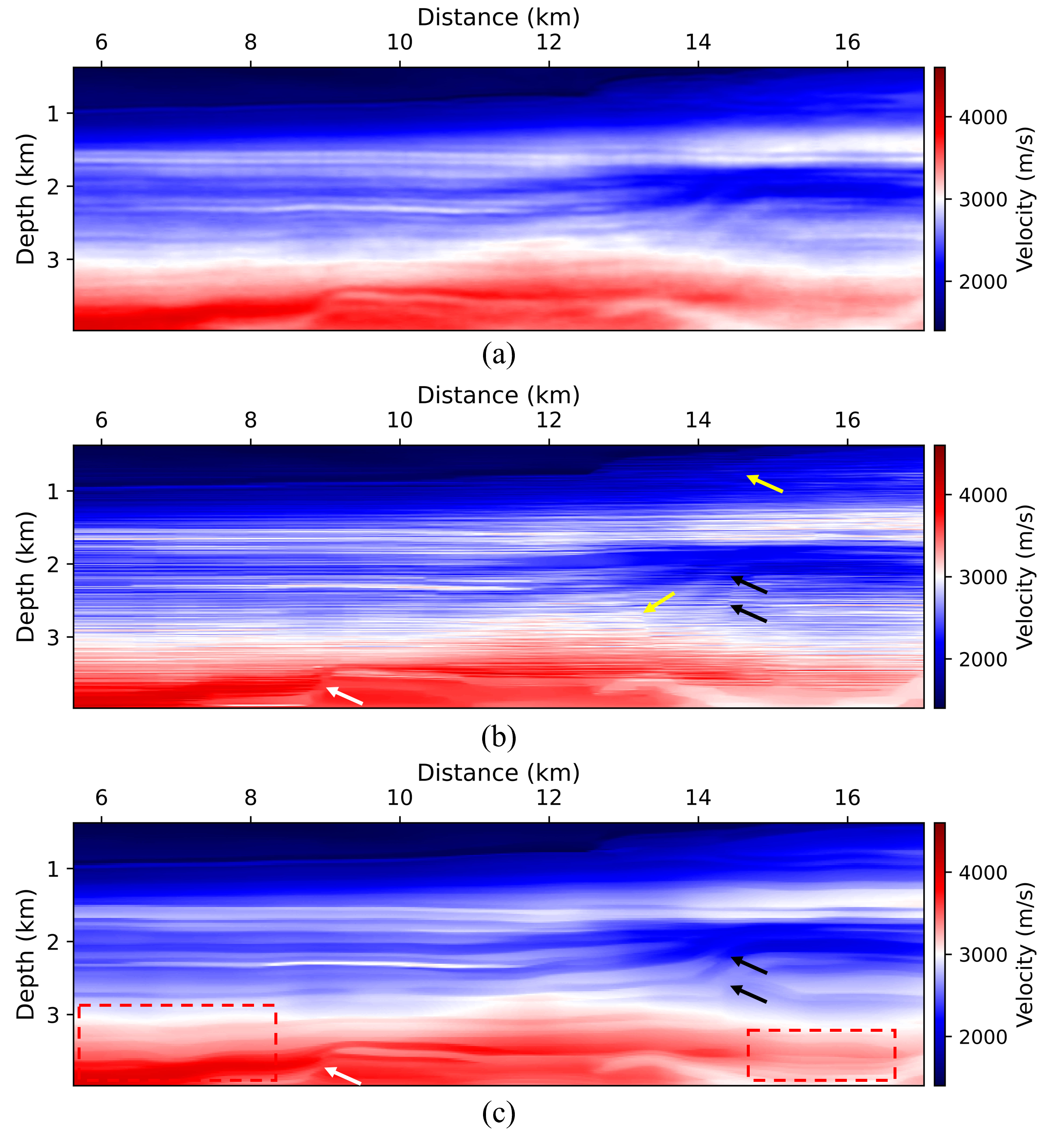}}
    
\caption{Prior injections for the CGG field dataset using the FWI result only. (a) the FWI result. (b) the prediction using the Otway priors (c) the prediction using the CGG priors }

\label{fig:cgg_results}
\end{figure}
\begin{figure}[htbp!]
    \centering
      {\includegraphics[width=0.8\columnwidth]{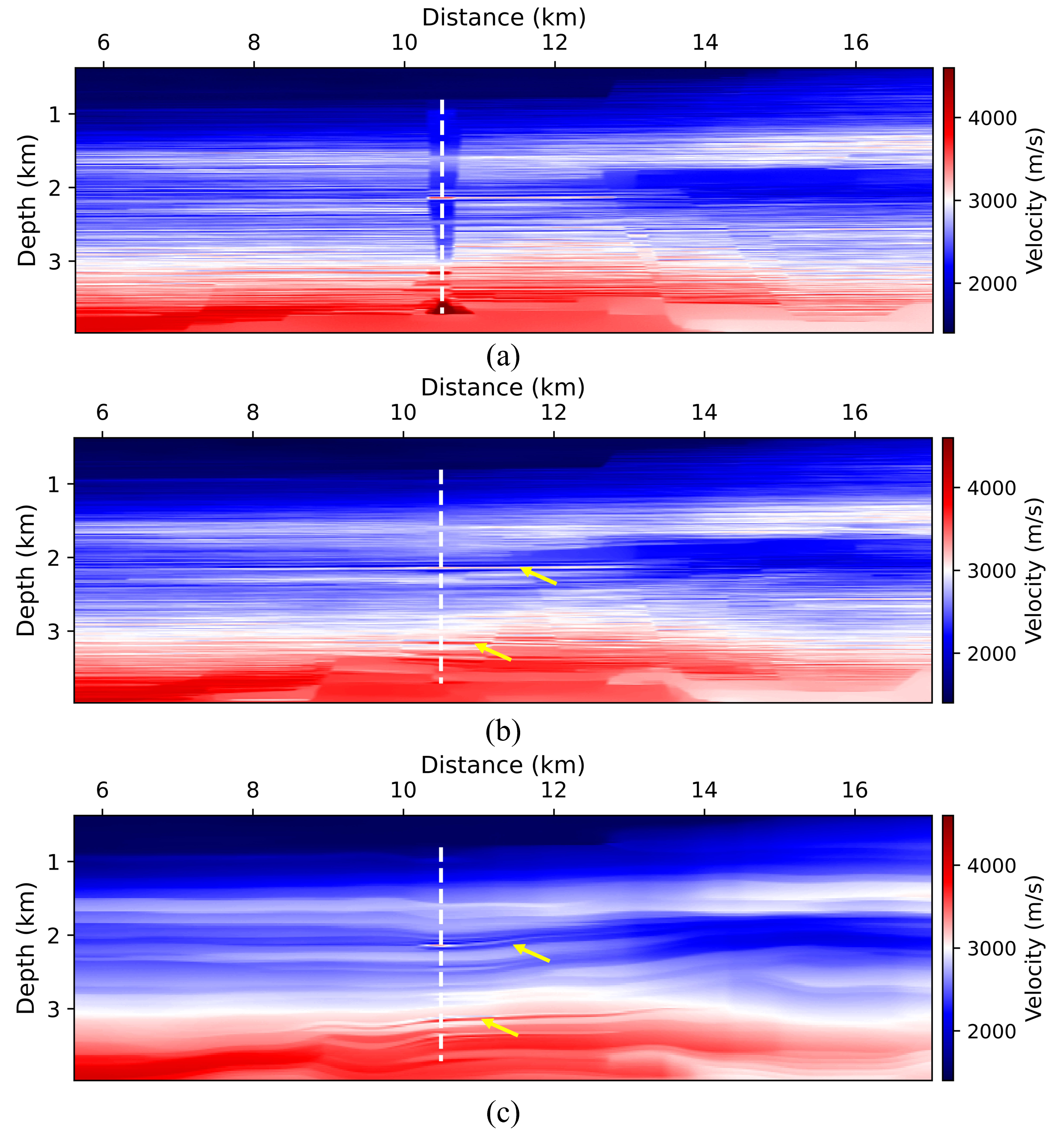}}
    
\caption{Prior injections for the CGG Field dataset using the FWI inverted result constraint and the well-log constraint. (a) the Otway priors using Equation \ref{eq:19}. (b) the Otway priors using Equation \ref{eq:20}. (c) the CGG priors using Equation \ref{eq:20}.}

\label{fig:cgg_welllog_results}
\end{figure}

\begin{figure}[htbp!]
    \centering
      {\includegraphics[width=1\columnwidth]{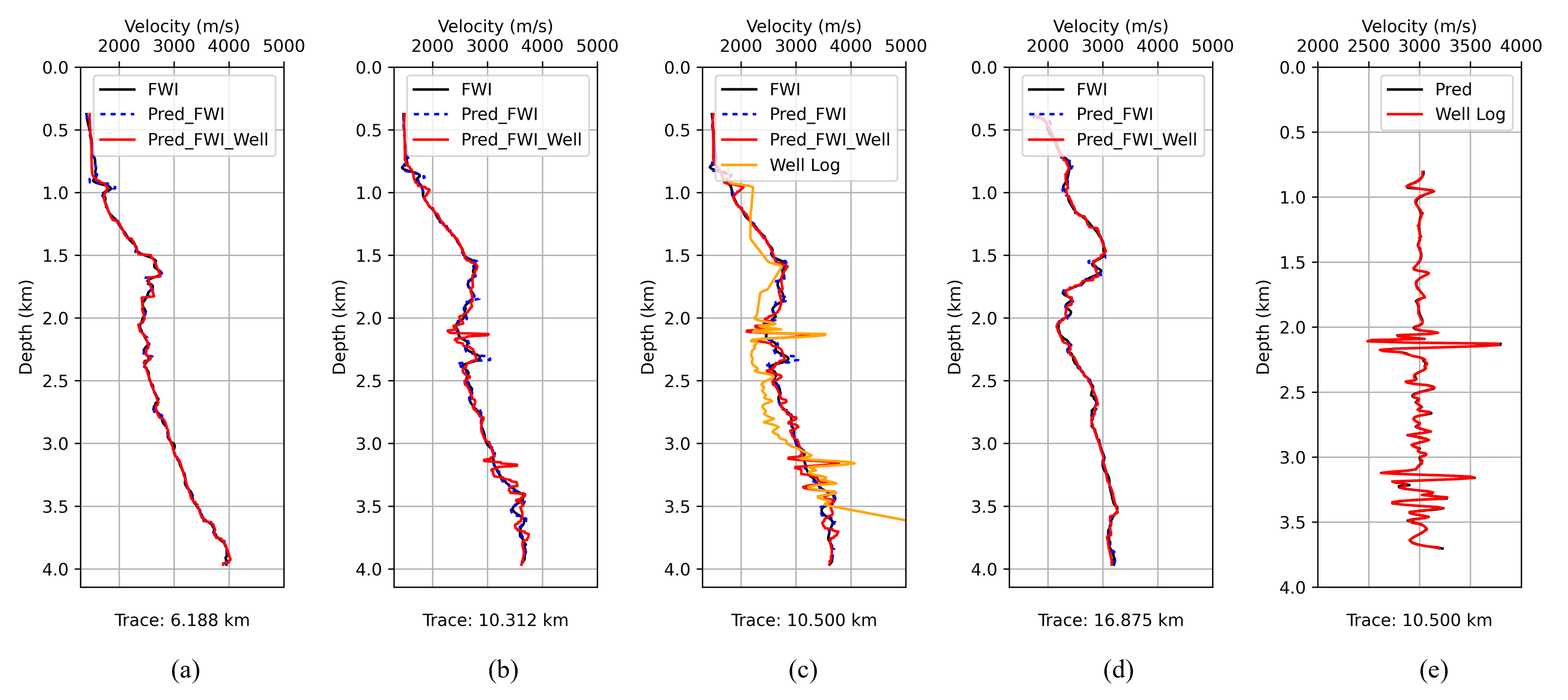}}
    
\caption{Comparison of velocity profiles of the predicted results for the CGG field example. (a), (b), and (d) show profiles located away from the well, while panel (c) corresponds to the well location. (e) presents the filtered high-frequency components of the predicted result and the well log.}
\label{fig:cgg_wellog_comparsion}
\end{figure}

\begin{figure}[h!]
    \centering
      {\includegraphics[width=1\columnwidth]{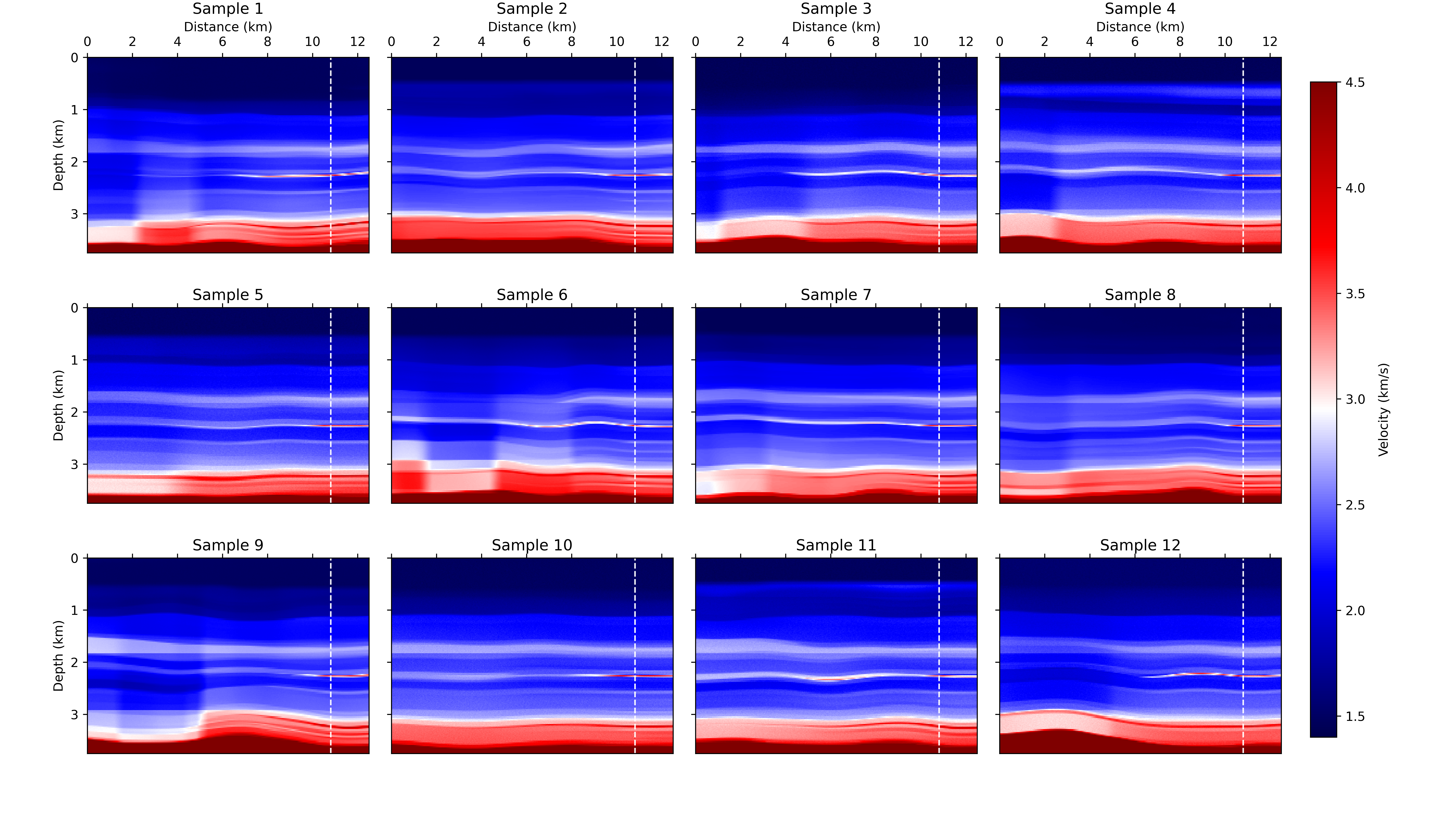}}
    
\caption{Twelve generated velocity models guided by a single well log. The dashed line indicates the well location.}

\label{fig:cgg_well_log_only}
\end{figure}

\subsection{Field-Data Example: CGG}
We further evaluate the proposed method on a CGG field dataset. The extracted velocity model extends from location \(5.6235\) km to location \(17.025\) km with \(608\) lateral sampling points. The depth range extends from \(0.375\) km to \(3.975\) km with \(608\) vertical sampling points. For generation, we set \(T=100\), \(t_s=30\), and \(N=1\). It is important to note that the spatial size of this dataset is significantly larger than that of the training samples used for the FM model. To initialize the generation process, the noise variable \(z\) is sampled from a Gaussian distribution with the same spatial dimensions as the target model, and Equation~\ref{eq:7} is used to construct the initial latent variable \(x_{t_s}\). Consequently, the U-Net directly processes the full-resolution field-scale input during inference. Notably, although the FM model is trained on relatively small size prior samples, it can still be directly applied to field-scale velocity models during inference without patch-wise reconstruction or resizing.

Figure \ref{fig:cgg_results} (a) shows the FWI result for this dataset, where noticeable inversion artifacts are present and the resolution is limited. Figures \ref{fig:cgg_results} (b) and \ref{fig:cgg_results} (c) present the results after injecting the Otway and CGG priors, respectively. When the Otway priors are used, thin layers are generated, and a fault at the bottom is recovered, as indicated by the white arrow. It also seems to suggest the presence of additional faults (shown by the yellow arrows), as the Otway priors promote such extensive faulting. When the CGG priors are injected, the layers become more laterally continuous in a geologically meaningful manner, and the resolution of layer boundaries is further improved (see red box area). In addition to the bottom fault, a fault located at the top right might also be present, as shown by the black arrows. 

Figure~\ref{fig:cgg_welllog_results} presents the results further constrained by the available well log located at 10.5 km. A straightforward optimization-based guidance strategy using Equation.~\ref{eq:19} can be applied to jointly constrain the generation using both the FWI result and the full-band well-log data. However, the significant mismatch between the smooth low-resolution FWI velocity and the high-resolution well-log velocity at the well locations makes the optimization unstable and difficult to converge when directly applying Equation.~\ref{eq:19}. As shown in Figure~\ref{fig:cgg_welllog_results} (a), this results in abrupt velocity changes around the well-log regions.

To alleviate this issue, Equation.~\ref{eq:20} is introduced to inject only the high-resolution components from the well logs while preserving the large-scale structures provided by the FWI result. The corresponding results using the Otway prior and CGG prior are shown in Figures~\ref{fig:cgg_welllog_results} (b) and (c), respectively. Compared to Figure~\ref{fig:cgg_welllog_results} (a), the abrupt velocity variations around the well locations are significantly reduced. As indicated by the yellow arrows, high-resolution layered structures are successfully reconstructed in both results, as indicated by the yellow arrows. In particular, compared to the otway prior result, the result generated using the CGG prior exhibits more geologically plausible layer continuations that naturally follow the large-scale structural trends defined by the FWI result.  This is because the geological prior plays a critical role in determining the spatial extension of the reconstructed high-frequency structures, as the well logs only provide sparse local constraints.

The effectiveness of the proposed method is further illustrated through the velocity profile comparisons shown in Figure~\ref{fig:cgg_wellog_comparsion}. Profiles (a), (b), and (d) are located away from the well location, while profile (c) is taken at the well location. We observe that the prediction with well constraints achieves a clear global improvement in resolution, even at positions far from the well. Figure~\ref{fig:cgg_wellog_comparsion} (e) represents the  high-frequency filtered
components of the predicted result and the well log. We can find that the predicted result successfully incorporates the high-wavenumber information from the well log. The difference in the low wavenumber is due to the difference between the vertical velocity (well velocity) and the normal moveout velocity (FWI velocity) due to the known anisotropy in the region \cite{sun2023anisotropic}.

To further demonstrate the role of well-log guidance, Figure~\ref{fig:cgg_well_log_only} presents generation samples constrained only by well logs using different random initializations of noise. We can see that the generated structures exhibit natural layer continuations and meaningful geological patterns learned from the priors, demonstrating that the diffusion prior can effectively infer globally consistent structures from sparse local well-log constraints.

\begin{figure}[h!]
    \centering
      {\includegraphics[width=1\columnwidth]{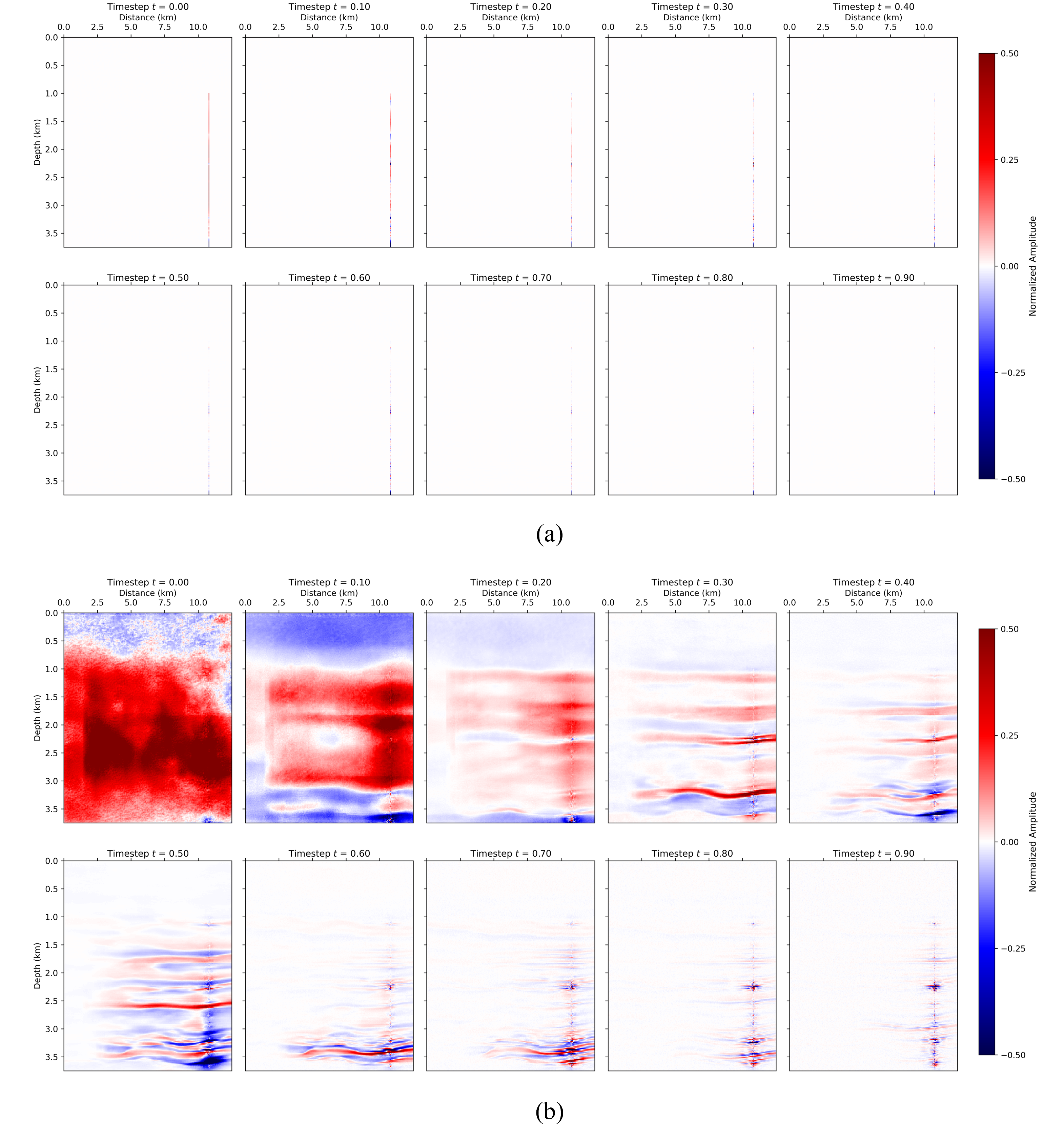}}
    
\caption{Comparison between the gradients from the masking forward operator $\nabla{\hat{x}_1}\mathcal{L}$ (a) and the gradients of the latent variables $\nabla{\hat{x}_t}\mathcal{L}$ (b). (b) is obtained by further backpropagation (a) through the U-Net. Directly applying the gradient in (a) to update the latent variables would fail, since nonzero gradients exist only at the well-log locations. In contrast, (b) exhibits a fully globalized update pattern extending away from the well-log location, because the gradients in (a) are backpropagated through the U-Net, and thus, benefits from the learned prior.}

\label{fig:gradient_comparasion}
\end{figure}

To better understand why the proposed well-log guidance strategy is effective, we visualize the gradients with respect to both \(\hat{x}_1\) and \(x_t\). Figure~\ref{fig:gradient_comparasion}(a) shows the gradient \(\nabla_{\hat{x}_1}\mathcal{L}\). Since a masking forward operator is applied to \(\hat{x}_1\), the resulting gradients are nonzero only at the well-log locations. For conventional methods, such highly localized gradients must be manually extrapolated to neighboring regions in order to influence the global model. However, this extrapolation process often depends heavily on human-designed assumptions and may fail in complex geological settings. In contrast, our method further backpropagates the guidance gradients through the U-Net prior, resulting in the latent-space gradient \(\nabla_{x_t}\mathcal{L}\), shown in Figure~\ref{fig:gradient_comparasion} (b). Unlike the masked local gradients in (a), the gradients in latent space are naturally distributed over a much broader spatial region. This indicates that the learned prior embedded in the network automatically propagates sparse local observational constraints into globally coherent structural updates.

\subsection{Field-Data Example: Viking}
\begin{figure}[h!]
    \centering
      {\includegraphics[width=1\columnwidth]{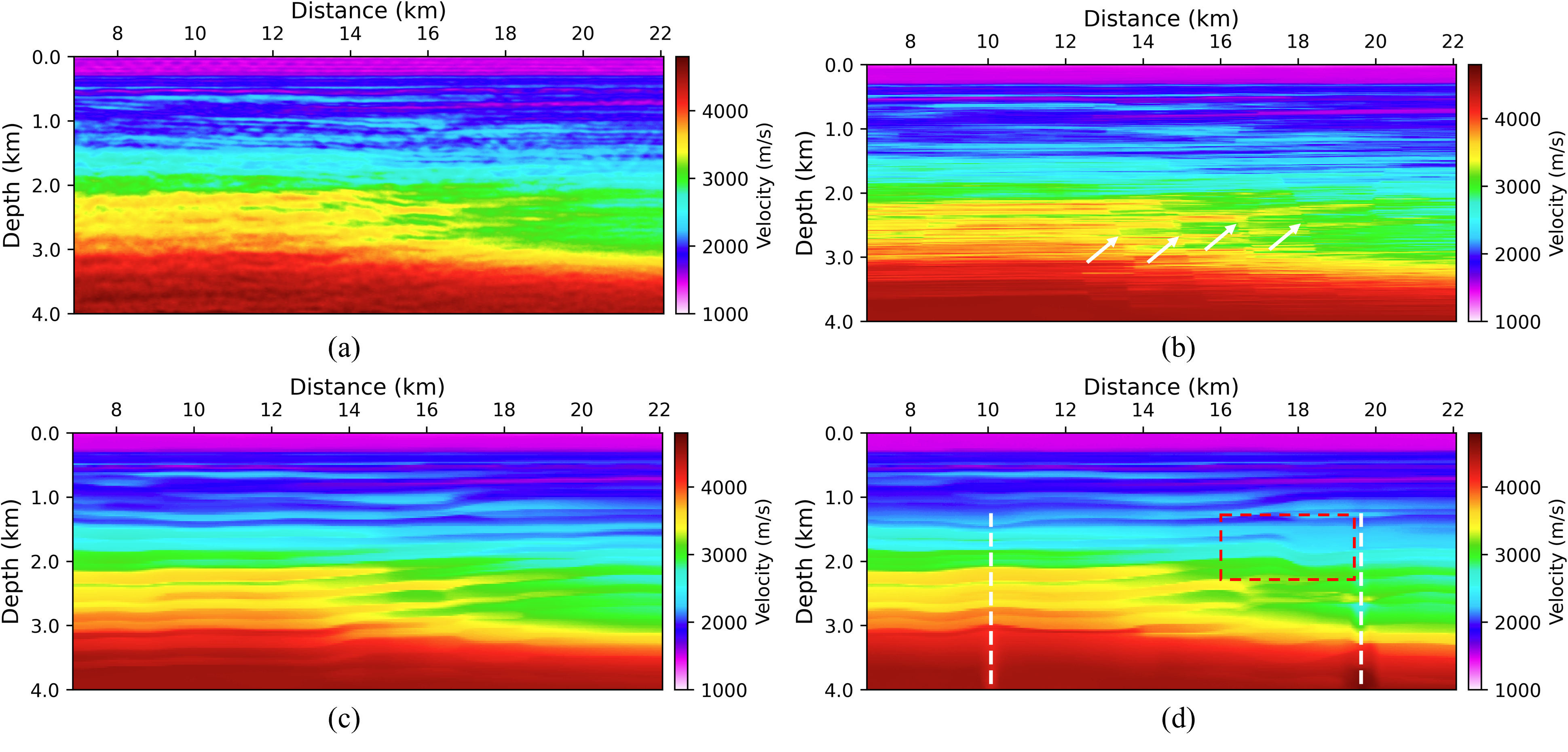}}
    
\caption{Prior injection results on the Viking field dataset. (a) FWI baseline result. (b) Prediction guided by the FWI result in (a) with Otway priors. (c) Prediction guided by the same FWI result using CGG priors. (d) Prediction guided by both the FWI result and well-log constraints using CGG priors. 
Note the divergence between (b) and (c), highlighting the impact of different priors. 
The two white dashed lines indicate the well locations and their available depth ranges.}

\label{fig:viking results}
\end{figure}

\begin{figure}[h!]
    \centering
      {\includegraphics[width=1\columnwidth]{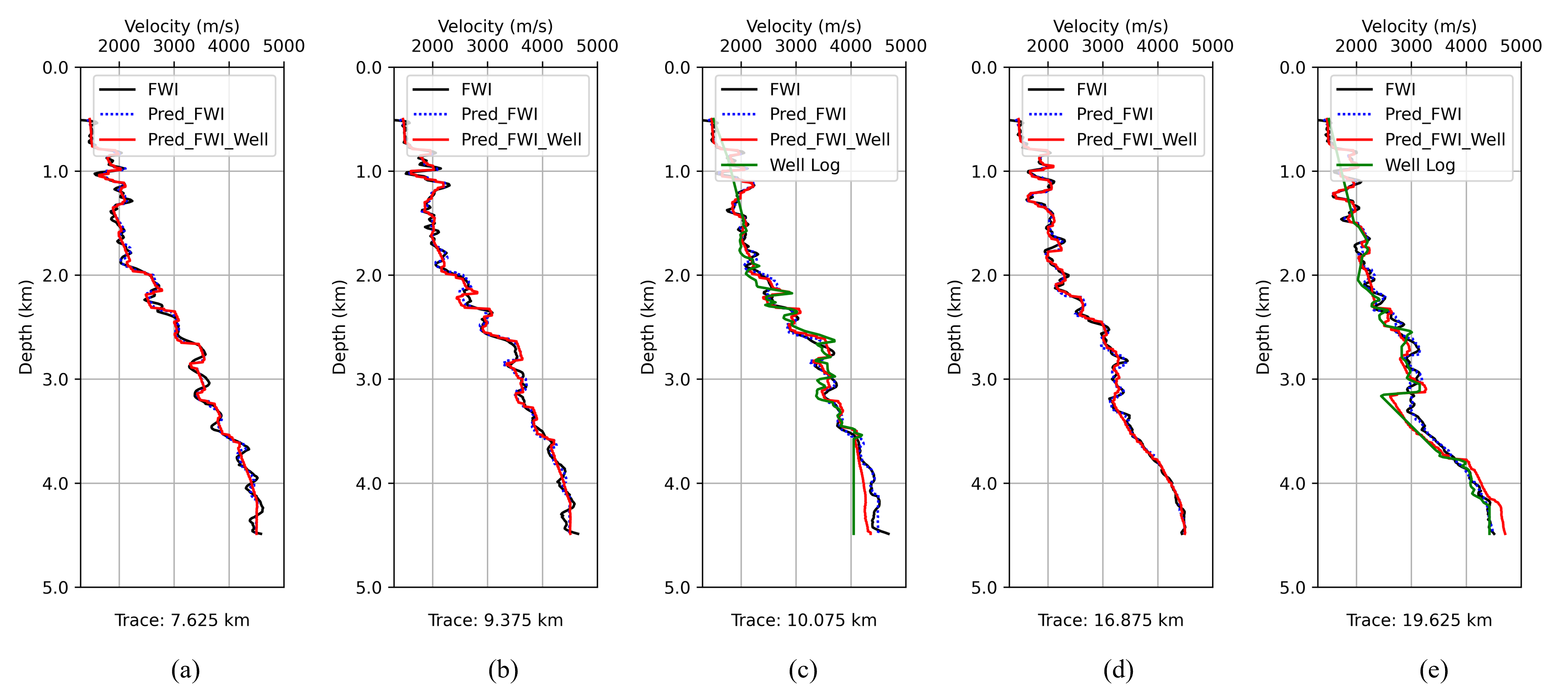}}
    
\caption{Comparison of velocity profiles of the predicted results for the Viking field example. (a), (b), and (d) show profiles located away from the well, while panel (c) and (e) correspond to the well location.}
\label{fig:viking_wellog_comparsion}

\label{fig:viking_well_comparasion}
\end{figure}

 Finally, we test the model on another field data, specifically, the Viking dataset. The maximum offset for this dataset is 3 km, which poses a challenge for FWI. Thus, while the FWI result in Figure \ref{fig:viking results} (a) recovers shallow structures, the deeper layers remain noisy due to limited illumination. For prior injection, we set \(T=100\), \(t_s=30\) and \(N=1\). The result obtained using the Otway prior injection is shown in Figure \ref{fig:viking results} (b). The noise is effectively suppressed; however, since the Otway priors mainly consist of thin horizontal layers and faults, the generated structures contain faults in regions where the layers should be more continuous (shown with the arrows). The result of CGG prior injection is shown in Figure \ref{fig:viking results} (c). In this case, the noise is largely removed and the layers are more consistent, showing a better match with the underlying geological structures. Figure \ref{fig:viking results} (d) shows the prediction result using the CGG prior with well-log constraints from Equation~\ref{eq:19}. The locations of the two provided well logs are indicated by white dashed lines, and their detailed profiles are presented in Figures~\ref{fig:viking_well_comparasion} (c) and~\ref{fig:viking_well_comparasion} (e). Since the well logs are largely interpolated between discrete measurements, they do not provide resolution higher than the FWI result in this case. The resulting well-log-constrained model does not show a continuous layer-wise constraint; however, as indicated by the red box, it still effectively guides the correction of local layer structures and improves the geological consistency of the result.

\section{What is next?}

Here, we shared an example of two types of prior that could be injected into the provided FWI result (for example, from a vendor) and those are our geological expectations and any well information. These geological expectations are represented by samples of Earth models we constructed to comply with what we expect the subsurface would look like in a specific area, including potential sharp layers, faults, folds, and other geological features. These samples are then used to train the FM model to store the (often referred to as prior) distribution of subsurface models from which these samples can be drawn. As a result, the training set for the FM model, as we saw, plays a major role in the type of prior we inject into the model. Far more advanced Earth model samples can be built using our geological knowledge, and those samples can be used to train the FM model. Injecting priors into the FWI inverted result is not limited to our geological expectations, but we can also inject well information and other data. In fact, the approach can be used to embed even high resolution information from, for example, near surface seismic data. We plan to test these capabilities in future work.

\section*{Conclusions}
We presented an FM based post-FWI refinement framework that leverages a learned generative model to improve FWI results through guidance towards the original FWI solution. Taking advantage of the deterministic nature of the FM ODE formulation, the proposed method enables effective integration of prior information while maintaining fidelity to the inverted FWI result. Synthetic and Field-data experiments demonstrate that the proposed approach effectively suppresses inversion artifacts and enhances resolution. These results highlight the potential of FM–based generative priors as a practical and efficient tool for post-FWI processing.

\section*{Acknowledgments}
This publication is based on work supported by King Abdullah University of Science and Technology (KAUST). The authors thank the DeepWave sponsors for their support. The authors would also like to thank Shuo Zhang and Linrong Wang for providing the FWI results used in this paper.

\bibliographystyle{unsrt}  
\bibliography{references}

\end{document}